\documentclass[twocol]{ametsocV6.1}

\title{
The Role of Deep Mesoscale Eddies in Ensemble Forecast Performance
}

\authors{Justin P. Cooke,\aff{a}\correspondingauthor{Justin P. Cooke, justin.cooke@uri.edu} 
Kathleen A. Donohue,\aff{a}
Clark D. Rowley,\aff{b}  
Prasad G. Thoppil,\aff{b} 
D. Randolph Watts,\aff{a} 
}
\affiliation{\aff{a}{Graduate School of Oceanography, University of Rhode Island, Narragansett, Rhode Island}\\
\aff{b}{Ocean Sciences Division, Naval Research Laboratory, Stennis Space Center}\\
}

\abstract{Present forecasting efforts rely on assimilation {}{techniques that adjust the model basic state, meaning that profiles of temperature and salinity are used as measured or converted to temperature and salinity through statistical relationships. This information influences} the upper ocean ( $< 1000$ m depth), {while} minimally {influencing the deep ocean}. Nevertheless, development of the full water column circulation critically depends upon the dynamical interactions between upper and deep fields. {A review of ensemble forecasts in the Gulf of Mexico demonstrates the importance of the initial deep ocean features in the evolution of the surface field. Initial conditions throughout the full water column that agree with observations are needed to improve the forecast predictions.} Here, best and worst ensemble members in two 92-day forecasts are identified and contrasted in order to determine how the deep ocean {features differ} between these groups. The forecasts cover the duration of the Loop Current Eddy Thor separation event, which coincides with available deep observations. Model member performance is assessed {with a newly developed ranking method, demonstrated with} surface variables against verifying analysis and satellite altimeter data during the forecast time-period. Deep cyclonic and anticyclonic features are reviewed, and compared against deep observations, indicating subtle differences in locations of deep eddies at relevant times. These results highlight both the importance of deep circulation dynamics of the Loop Current system and more broadly motivate efforts to assimilate deep observations to better constrain the deep initial fields and improve surface {and sub-surface} predictions.}

\begin{document}

\maketitle

%
%
%
%
%
%

%





\section{Introduction}

\begin{figure}[hbt!]
    \centering
    \includegraphics[width=0.85\linewidth]{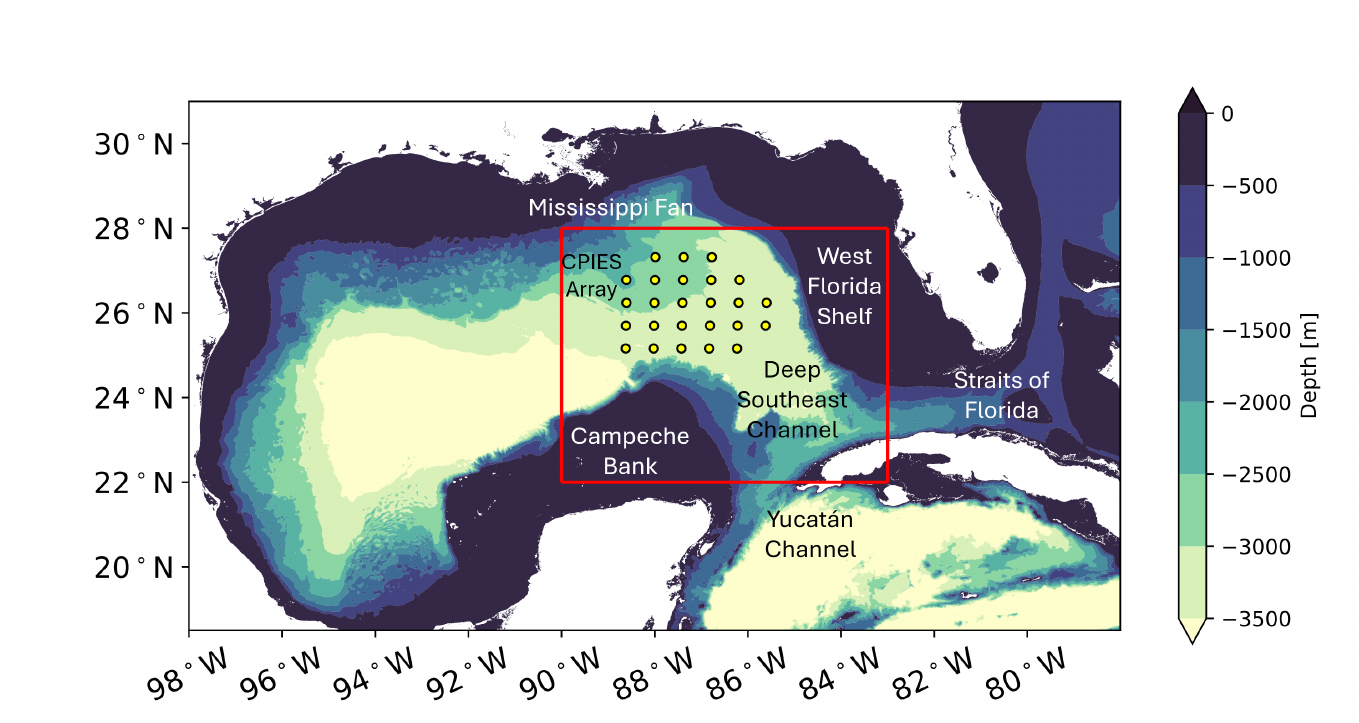}
    \caption{The Gulf of Mexico with the region of interest (red box), which ranges from $22^\circ$N to $28^\circ$N, $90^\circ$W to $83^\circ$W. 
    The Understanding Gulf Ocean Systems CPIES array (black, yellow-filled circles), and names of key {locations} are included. Bathymetry is contoured every $500$ meters.}
    \label{fig:gom}
\end{figure}

The Loop Current (LC) system (Fig.~\ref{fig:gom}) is the primary forcing mechanism for mesoscale oceanographic variability in the Gulf of Mexico~({GoM}, \citealt{oey2008loop}). 
The LC brings warm, salty waters from the Caribbean Sea \textit{via} the Yucat\'an Channel, exiting through the Straits of Florida. 
Flow may enter and exit in a retracted port-to-port mode, or it may extend into the Gulf as a northward loop as far as 28$^\circ$N and 93$^\circ$W. 
As the LC extends, it eventually pinches off a closed warm core, anticyclonic eddy with horizontal scales of 200-400 kilometers, known as a Loop Current Eddy (LCE).
The LCE may reattach or shed, advecting west, with an observed shedding frequency ranging from weeks to months~(\citealt{sturges2000frequency,leben2005altimeter}).
Before its eventual separation, the LCE may undergo multiple instances of detachment and reattachment~(\citealt{leben2005altimeter,tsei2025low}). 
LCEs carry significant energy and strong currents, affecting transport of heat and momentum, with implications for hurricanes~(\citealt{le2021role}), fisheries~(\citealt{cornic2018influence,gilmartin2020seasonal}) and offshore energy industries such as oil and gas~(\citealt{national2018understanding}).
By good fortune, during the 2010 \textit{Deepwater Horizon} oil spill, peripheral eddies generated by the LC were observed to have entrapped some of the oil, preventing it from continuing through the Straits of Florida~(\citealt{hiron2020evidence}).
Improving our understanding of the LC system, and the mechanisms contributing to LCE separation, is critical for ensuring both the safety of the communities around the Gulf, and for responsible operation by industries in the region.  

Numerous studies -- ranging from numerical~(\citealt{hurlburt1982dynamics,oey2008loop,chang2011loop,le2012simulating,dukhovskoy2015characterization}), observational~(\citealt{candela2002potential,donohue2016loop,sheinbaum2016structure}), and theoretical~(\citealt{pichevin1997momentum,nof2005momentum}), among many others -- have investigated the physics of LCE separation. 
Availability of altimeter derived surface measurements, along with an abundance of instruments operating in the upper 1000 meters, has resulted in a focus on {the impact of} upper oceanic processes~(\citealt{schmitz2005cyclones,sheinbaum2016structure,androulidakis2021impact,laxenaire2023impact}).
{Among many suggested mechanisms for LCE separations,} a prominent theme {are eddies on the periphery of the LC, known as} Loop Current Frontal Eddies (LCFEs){, which are believed to} contribute significantly to the eddy shedding process~(\citealt{schmitz2005cyclones,le2012simulating}).
These eddies {have} horizontal scales $\sim O(10-100)$ kilometers, reaching depths of $1000$ meters.
{A number of formation mechanisms have been suggested including LC interaction with topography, such as} near the Campeche Bank{, the Mississippi Fan, or the West Florida Shelf}~(see Fig.~\ref{fig:gom};~\citealt{oey2008loop,le2012simulating,sheinbaum2016structure}).
However, prior work has also demonstrated the existence of deep, mesoscale eddies in the Gulf~(\citealt{oey2008loop,hamilton2009topo,donohue2016loop,hamilton2016loop,perez2018dominant,johnson2022generation,safaie2025deep}).
These eddies exist below 1000 meters and interact with the upper baroclinic jet through upper-deep vortex stretching, related to baroclinic instability~(\citealt{hamilton2009topo,le2012simulating,donohue2016loop}), but their role in LCE shedding is still not  fully understood~(\citealt{hamilton2019loop}). 
Increased interest in the deep currents of the Eastern Gulf led to a recent investigation from~\cite{morey2020assessment}, who highlighted differences between deep observations and models: notably that contemporary models vastly underestimate deep eddy kinetic energy {(EKE)} levels generated during the LCE separation process.
A study from~\cite{rosburg2016three} demonstrated that models tend to underestimate the {EKE} in the deep by a factor of three. 
Nevertheless, despite these discrepancies in modeling deep energy, forecasts appear to reasonably represent the LC system~(\citealt{wei2016performance,thoppil2025evaluating}). 

Numerical models are used to forecast the ocean with Data Assimilation (DA), where {observations are assimilated to produce the analysis, which the forecasts begin from}~(\citealt{stammer2016ocean}). 
Forecasts can be run in single deterministic configurations, or as an Ensemble Forecast (EF) with multiple members that represent the best guess for a predictive state of a dynamical system~(\citealt{wei2008initial}).
Recent work from~\cite{thoppil2021ensemble}{, using forecasts with a 3-D Variational (3DVar) assimilation scheme,} demonstrated improved forecast skill -- by nearly a factor of three -- of ocean mesoscales for a lower resolution EF, compared against a higher resolution deterministic forecast. 
Forecast members may differ within the range of their error covariance -- the matrix composed of the variance of individual errors, and how they are related to other errors in the system -- which accounts for {model} uncertainty in the initial conditions~(\citealt{martin2015status}). 
EFs assimilate surface data from satellites, such as sea surface height (SSH) and sea surface temperature (SST), {and subsurface observations, such as} temperature and salinity profiles from \textit{in situ} observational platforms -- ARGO floats, CTDs, and gliders~(\citealt{rosburg2016three,wei2016performance}). 
A recent study from~\cite{dukhovskoy2023assessment} {reviewed the added value of including temperature and salinity profiles, found using the Gravest Empirical Mode} (\citealt{donohue2016gulf}){, from an array of Current and Pressure Inverted Echo Sounder (CPIES) data. The addition of the CPIES profiles, along with other profiles derived form surface and sub-surface observations} {substantially} improved model performance in the Gulf. 

Here, we demonstrate the impact of the unconstrained deep field on {EF} performance {in the GoM.
In the deep Gulf, two vertical modes constitute $\geq 90\%$ of the variance}~(\citealt{hamilton2016loop}){: a surface-intensified mode and a depth-independent mode.
Most of the vertical structure in the surface-intensified mode appears in the upper 1000 meters -- hydrographic properties have weak lateral changes below $1000$ meters. 
We adopt a simple terminology: ``upper'' refers to the surface-intensified mode and steric component of SSH; ``deep'' describes the nearly depth-independent mode. 
Our study assesses (i) the influence of the deep initial field on ensemble forecasts, and (ii) the spread of deep-field predictions within an ensemble forecast.}

{We focus on the time-period} during the LCE Thor Separation event, which spanned October 28th, 2019 through April 6th, 2020.
This event was challenging from a  forecast perspective, as LCE Thor detached, and eight weeks later reattached and then separated.
An analysis of eleven weekly ensemble forecasts between October 28th, 2019 through January 6th, 2020 reveal that the initial conditions that agree better with deep observations produce improved upper {ocean} forecasts 5 and 6 weeks later. 
Then proceeding from this finding, we examine the coupled upper-deep processes developing during two EF periods, one beginning October 28th, 2019 and the other on January 6th, 2020, described in greater detail in~\cite{thoppil2025evaluating}{. These forecasts are selected due to their efficacy} in {accurately} predicting the complex separation process.
We implement a strategy to identify the best and worst performing ensemble members, based on SSH, using two different benchmarks to compare method robustness.
The deep fields for the two groups are contrasted to identify the differences leading to their predictions.
Moreover, mapped observations of the deep fields from CPIES are available during this time-period, allowing for direct comparison.
Our work highlights how deep observations could be effectively used by regional ensemble forecasts in strong current regimes to improve forecasts.

\section{Methodology}

\subsection{Ensemble Forecast Model}

The Naval Research Laboratory runs a 32-member Ensemble Forecast system in the Gulf of Mexico. 
This forecast system combines the Navy Coastal Ocean Model~(NCOM;~\citealt{martin2009user}) with the Navy Coupled Ocean Data Assimilation system {for DA} (NCODA;~\citealt{cummings2005operational}), as well as {forcing from} the Coupled Ocean-Atmosphere Mesoscale Prediction System (COAMPS;~\citealt{hodur1997naval}).
{Further details of the model are found in Appendix C.}
The domain includes the entire Gulf of Mexico (Fig.~\ref{fig:gom}), and has 3 km horizontal resolution and 49 hybrid levels in the vertical.
The vertical levels are split into 33 sigma (terrain/surface following) levels and below that 16 z-, or pressure, levels; the finest resolution is near the surface, becoming coarser with depth. 
Vertical mixing is parametrized with the Mellor-Yamada level 2 turbulence closure scheme~(\citealt{mellor1982development}). 
Lateral boundary conditions are handled as tides{. 
Tidal forcing is applied to the domain, and tidal open boundary conditions for water level and barotropic velocity are provided by the Oregon State University global Ocean Tide Inverse Solution(OTIS;}~\citealt{egbert2002efficient}{). 
Thus, locally generated internal tides are present in the model, more information is included in Appendix C.
The forecast is begun from an NCODA analysis product, which assimilates all available observations.} 

The NCODA system deploys a {3DVar} assimilation scheme, allowing for near real-time inclusion of observations, such as satellite altimeter products (SSH and SST), as well as {temperature and salinity} profiles from gliders, floats, and ships.
Satellite altimeter observations are incorporated into the model \textit{via} the Modular Ocean Data Assimilation System (MODAS); MODAS synthesizes temperature and salinity profiles from SSH and SST, projecting them down through the water column to constrain the ocean interior~(\citealt{fox2002modular}). 
Velocity information is not assimilated into the forecast system. 

In order to account for a variety of physically realizable and dynamically relevant ocean states, model initial conditions are perturbed. 
The perturbations are formulated from the 24-hour forecast error variances of a control run, which represent model and observation uncertainties in a single deterministic forecast, using the Ensemble Transform method~(\citealt{bishop1999ensemble,wei2006ensemble}).
Additionally, estimates of the model temporal variability and climate variability are added to ensure a similar spread of the ensemble model perturbations to the error variance of the best guess control run. 

Ensemble forecasts from two starting dates are analyzed, chosen to span the entirety of the LCE Thor separation event. 
The focus region of the study is the Eastern Gulf, where the LCE separation occurs, bounded by $22^\circ$N to $28^\circ$N, $83^\circ$W to $90^\circ$W (red box in Fig.~\ref{fig:gom}).
The first 92-day ensemble forecast begins on October 28th, 2019, and ends January 27th, 2020, and is referred to as EF20191028. 
This forecast successfully predicted the timing of the initial detachment of the LCE, occurring around January 10th, coinciding with the presence of a cyclonic eddy in the region of the Deep Southeast Channel (DSC; Fig.~\ref{fig:gom}).
The second forecast begins on January 6th, 2020, and ends April 6th, 2020, and is referred to as EF20200106. 
This time-period covers the observed reattachment of the LCE near March 16th, and final separation around March 30th, associated with a deep cyclone moving south and westward off the Mississippi Fan (Fig.~\ref{fig:gom}).

\subsection{Observations}

Two datasets{, which were not included in the observations assimilated into the Ensemble Forecasts,} are used to assess model member performance:  an array of Current Pressure Inverted Echo Sounders (CPIES) and satellite sea surface height derived from~\cite{cmemsdata}. 
Twenty-four CPIES (black, yellow-filled circles in Fig.~\ref{fig:gom}) were deployed for almost two years, June 2019 through May 2021 and span the LCE Thor separation event. 
The array extended from $25^\circ$N to $27.5^\circ$N, $86^\circ$W to $89^\circ$W, with approximately $60$ km spacing between instruments~(\citealt{johnson2022generation}).  
Here we focus on the near-bottom pressure and current CPIES records.  
Details regarding CPIES processing can be found in \citealt{johnson2022generation}.  Pressure and currents are low-pass filtered with a 72-hour cutoff, and subsampled in 12-hour increments.  
Bottom pressure maps are generated using a multivariate optimal interpolation approach, which constrains pressure and velocity to be geostrophic{. Including} deep velocity observations sharpens the gradients~(\citealt{watts2001mapping}). 

For comparisons to estimate the accuracy of the forecast SSH fields, we use global CMEMS satellite data from October 28th, 2019 to April 6th, 2020 from $22^\circ$N to $28^\circ$N, $80^\circ$W to $90^\circ$W.
The data consists of Absolute Dynamic Topography (ADT) which is daily averaged and gridded with $0.125^\circ\times0.125^\circ$ horizontal resolution{, and effective resolution $\sim{O(100)}$ km}~(\citealt{ballarotta2019resolutions}){. Despite the coarse effective resolution, the scales of the LC and LCE $\sim{O(200-400)}$ km are still adequately resolved}.
For consistency in the analysis, we remove daily spatial means {(within our region of interest; red rectangle in Fig.~\ref{fig:gom})} from the SSH fields, such that the 17-cm contour line, representative of the LC and LCE fronts~(\citealt{leben2005altimeter}), may be tracked and used for comparison against the model. 

\subsection{Identification of Deep Mesoscale Eddies}

To represent deep circulation and mesoscale eddies in the model data, we calculate the deep reference pressure by expressing the total sea surface height ($\eta_{total}$ [m]) as the sum of the reference ($\eta_{ref}$ [m]) and steric {($\eta_{steric}$ [m])} components. 
Use of $\eta_{ref}$ is advantageous in its simplicity -- eddies are identified using this scalar stream function, as opposed to a more highly differentiated scalar parameter like relative vorticity.
This stream function enables a straightforward visualization of the deep fields, allowing for the determination of deep, mesoscale features. 

\begin{equation}
    \eta_{total} = \eta_{ref} + \eta_{steric}.
\end{equation}

The reference height may be converted to a reference pressure with units of Pascals (Pa), $\eta_{ref} \equiv P_{ref}/(\rho_b{g})$, where $\rho_b$ is bottom density and $g$ is gravity.  
The steric component is found by the geopotential height anomaly ($\phi$), divided by the acceleration due to gravity.
We can determine $\phi$ by integrating from the surface ($P_1 = 0$ dbar) to a constant reference pressure, $P_2 \approx 2023$ dbar (equivalently $2000$ m depth),

\begin{equation}
    \phi = \int_{P1}^{P_2} \delta dP,
\end{equation}

\noindent where $\delta$ is the specific volume anomaly,

\begin{equation}
    \delta = \frac{1}{\rho(S,T,P)} - \frac{1}{\rho(S_0,T_0,P)}.
\end{equation}

\noindent Here, $\rho$ is the density, $S$ is salinity, and $T$ is temperature, with reference values $S_0 = 35$ psu and $T_0 = 0^\circ$C.

\section{The Deep Field and SSH Prediction}

Deep mesoscale eddies can interact with the upper layer through vortex stretching. 
Upper and deep layer interaction may occur in two ways.
First, deep eddies are depth-independent, maintaining a uniform vertical structure throughout the water column; their associated reference velocities can add normal components crossing the upper layer baroclinic currents~(\citealt{donohue2016loop}).
Second, this interaction produces vertical stretching by the deep layer exerted on the upper. 
In this interaction the deep layer influences the upper layer, when its deep reference current has a component perpendicular to the slope of the upper layer isopycnals.
Thus its depth-independent, nearly isopycnal component produces upper layer vertical stretching. 
Conversely, the upper layer exerts influence on the deep when upper baroclinic fronts shift laterally and the sloped isopycnals produce vertical stretching of the weakly stratified lower layer. 
These two mechanisms occur together in baroclinic instability for instances where there is a favorable vertical offset of perturbations between upper and deep eddy variability.

\begin{figure}[hbt!]
    \centering
    \includegraphics[width=0.85\linewidth]{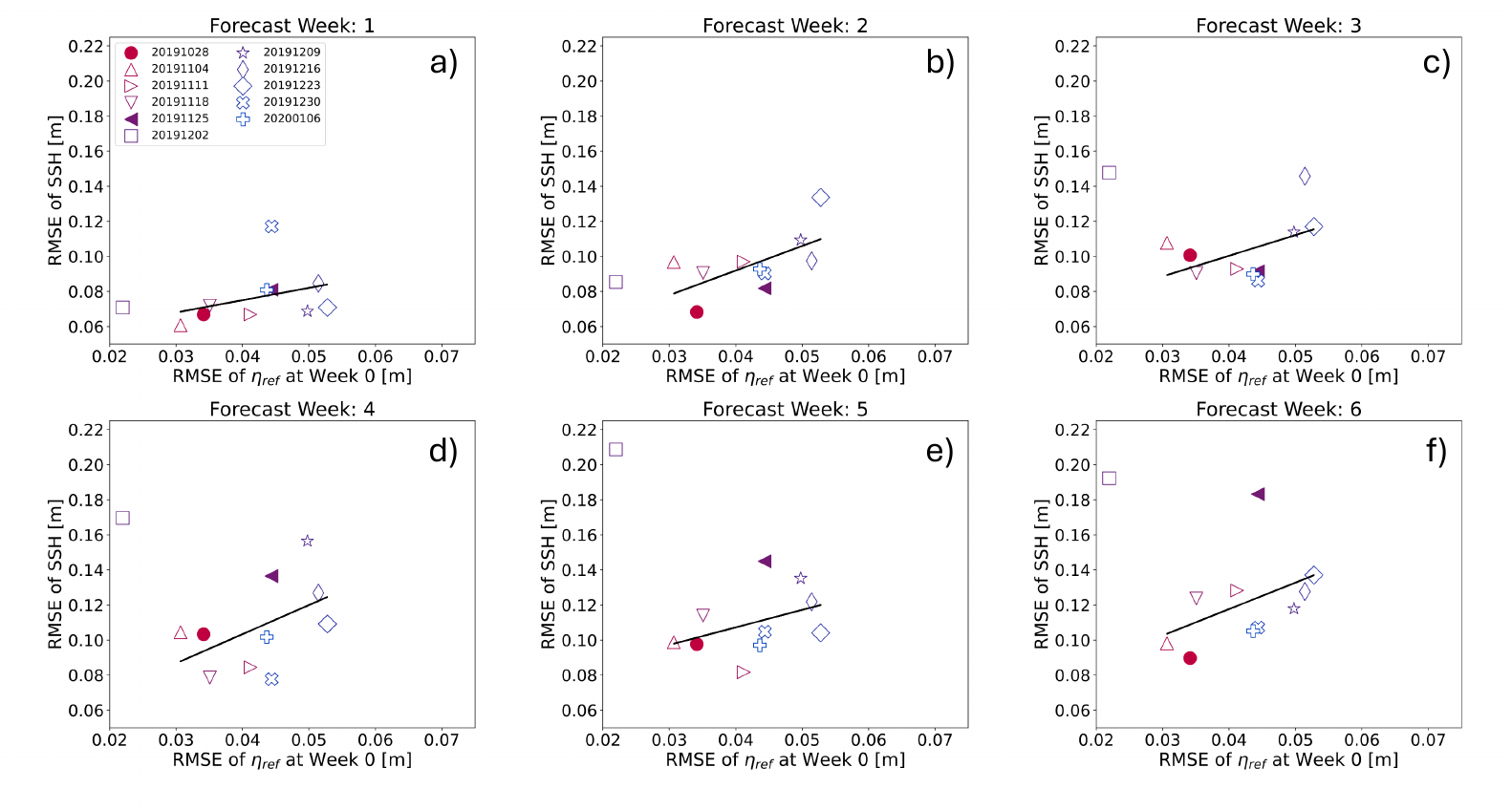}
    \caption{Scatter plot of RMSE of $\eta_{ref}$ from different forecasts at Week 0 compared against the observations from the CPIES array, plot against RMSE of SSH for each forecast's respective a) Week 1, b) Week 2, c) Week 3, d) Week 4, e) Week 5, and f) Week 6. {Linear trends (black, solid lines) are provided with $R^2$ values ranging from $0.11-0.38$.} Deep mean fields for $\eta_{ref}$ are shown in Fig.~\ref{fig:IC_vs_wk6} for the filled symbols.
    }
    \label{fig:rmse_etaref_vs_ssh}
\end{figure}

The impact of uncertainty in the deep field to forecast success is highlighted in (Fig.~\ref{fig:rmse_etaref_vs_ssh}) which shows the relationship between the initial, Week 0, root-mean-square-error (RMSE) $\eta_{ref}$ against weekly SSH RMSE.  Here, we use the  verifying analysis as {our ground truth} because it  constitutes the initial ocean state for each subsequent weekly long-range ensemble forecast and assimilates observations; therefore, it  represents the most realistic ocean state.
For $\eta_{ref}$ RMSE, we used Week 0 forecast ensemble mean and the observed CPIES $\eta_{ref}$ in the region $22^\circ$N to $28^\circ$N, $90^\circ$W to $83^\circ$W (see Fig.~\ref{fig:gom}).
Eleven weekly 92-day, 32-member ensemble forecasts during the LCE Thor time-period, from October 28th, 2019 through January 6th, 2020 are used in the analysis. 
Forecasts are referred to in the following manner: EFYYYYMMDD, where the date is the start of the forecast, \textit{i.e.,} October 28th, 2019 is EF20191028.
Three notable results emerge: 

\begin{enumerate}
    \item[i)] Figure~\ref{fig:rmse_etaref_vs_ssh}a indicates {the forecasts show} the initial uncertainty of the deep $\eta_{ref}$ field {weakly depends} on the stage of the LC cycle{. At Week 0, we expect RMSE of $\eta_{ref}$ to be minimal. E}arlier forecasts begin with a more retracted LC, with other forecasts beginning with an LC extended further into the Gulf, or a detached LCE.
    \item[ii)] Better initial conditions in the deep $\eta_{ref}$ field yield lower uncertainty in the upper SSH field as the forecast progresses, as indicated by the linear trend ({black, solid lines;} Fig.~\ref{fig:rmse_etaref_vs_ssh}).
    \item[iii)] By the sixth forecast week, a strong dependence is evident of the outcome upon the initialization. The members that have highest uncertainty in SSH, are approximately six weeks out from the initial detachment of LCE Thor, exemplified by two consecutive forecasts, EF20191125 (left triangle) and EF20191202 (square), which show highest uncertainty at Week 6 (Fig.~\ref{fig:rmse_etaref_vs_ssh}f).
\end{enumerate}

\begin{figure}[hbt!]
    \centering
    \includegraphics[width=\linewidth]{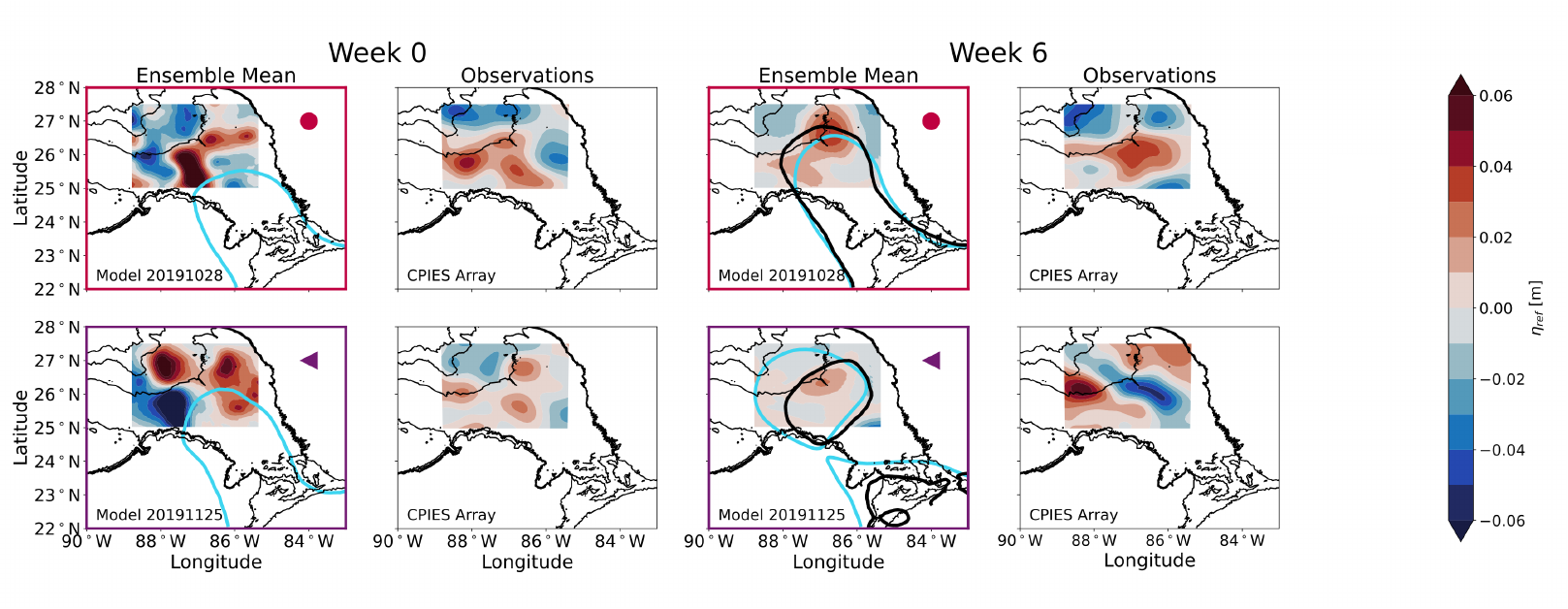}
    \caption{Deep ensemble mean $\eta_{ref}$ fields for EF20191028 (top row) and EF20191125 (bottom row), exemplifying a ``better'' initial condition and a ``poor'' initial condition, respectively.
    Model fields are contrasted against the CPIES observations at Week 0 (left columns) and Week 6 (right columns). 
    The {cyan} lines represent the 17-cm contour for the forecast, and the black lines at Week 6 represent the 17-cm contour for the verifying analysis.
    }
    \label{fig:IC_vs_wk6}
\end{figure}

To highlight the difference between a ``better'' and ``poor'' initial deep field, we plot the ensemble mean $\eta_{ref}$ field at 2000 meters for EF20191028 (top row in Fig.~\ref{fig:IC_vs_wk6}) and EF20191125 (bottom row in Fig.~\ref{fig:IC_vs_wk6}), these correspond to the two filled symbols in Fig.~\ref{fig:rmse_etaref_vs_ssh} {(circle and left triangle, respectively)}.
Model $\eta_{ref}$ field at Week 0 and Week 6 is contrasted with the CPIES observations, and the mean 17-cm contour of the forecast (cyan line) against the verifying analysis 17-cm contour (black line). 
Between the two forecasts, the initial field for EF20191028 contains more similar features as seen in the observations, including the more northern cyclone coupled with an anticyclonic feature to its south. 
Whereas the initial field for EF20191125 shows a strong anticyclone-cyclone dipole at $88^\circ$W that is not present in the observations, and {displays} a stronger anticyclonic feature {near $86^\circ$W than what is observed by the} CPIES.
{Any agreement between the deep initial field of the forecast and the observations is unexpected, as no information of the deep features are assimilated into the forecasts.}
As the forecast evolves, these initial conditions result in a better predicted deep field at Week 6 for EF20191028{; despite the model and observations disagreeing with the positioning, both display} an anticyclone {of similar magnitude} off the Mississippi Fan{, highlighting that the forecast is capturing a key feature of the deep ocean circulation}.
This is also reflected in the good agreement between the predicted 17-cm contour (cyan line) of the forecast, and the 17-cm contour of the verifying analysis on that date (black line).
In contrast, the deep field of EF20191125 shares no similarities with the CPIES observations{. In the SSH, the forecast appears to have only just detached, resulting in a more extended LC,} and shows a larger LCE. 
These findings imply a connection between deep skill and upper performance.

\section{Assessment of Ensemble Forecast Member Performance}

Two forecasts {are chosen as case studies, based on their high predictive skill during LCE Thor's separation}~(\citealt{thoppil2025evaluating}), to illustrate the relationship between upper and deep circulation. 
The first forecast, EF20191028, predicted the initial detachment of LCE Thor, and the second forecast, EF20200106, captured the reattachment and final separation of LCE Thor. 
As a first step,  best and worst performing members are isolated. 
The selection process is presented in the next section.

\subsection{Best and Worst Member Selection}

\begin{figure}[hbt!]
    \centering
    \includegraphics[width=0.75\linewidth]{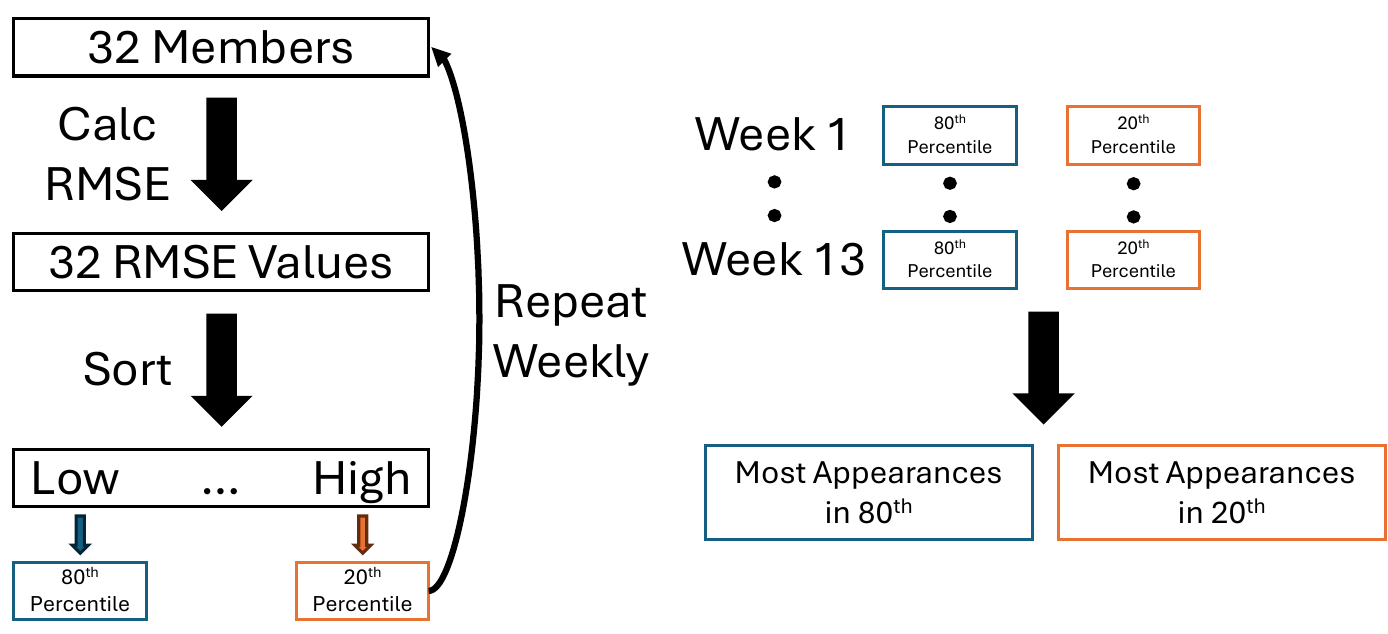}
    \caption{Schematic flow chart for member selection.}
    \label{fig:memSelect}
\end{figure}

Figure~\ref{fig:memSelect} outlines the procedure for selecting the best and worst members.
Across all members, we calculate the RMSE of a parameter of interest between the forecast and our chosen benchmark, at weekly intervals.
We sort members from lowest to highest RMSE, and use the $80^{th}$ and $20^{th}$ percentiles as cutoffs; the members within lowest and highest percentiles are recorded by week.
The number of appearances {over the thirteen weeks} for each member in each percentile is recorded, sorting from most to {fewest appearances}. For each group, {we select} the upper $80^{th}$ percentile{, i.e., those with the most appearances, determining} the best and worst members over the whole forecast time-period.
This selection process is generalizable to any desired variable.

This study focuses on comparing SSH, due to forecasts only assimilating data from the upper ocean. 
We use two ground truths: i) the verifying analysis; and ii) a gridded satellite product.
Note, we alternatively tested using surface velocity, but because we found a linear relationship between SSH RMSE and surface velocity RMSE, we dismissed the latter parameter (see Appendix A).
Additionally, we investigated using a method with information in the upper $500$ meters of the water column, composed of SSH, temperature ($T$), salinity ($S$), zonal velocity ($U$), and meridional velocity ($V$) from the verifying analysis.
A weighted RMSE value was found at each depth, and used for ranking. 
This method selected a majority of the same members -- for both the best and worst groups -- for each forecast, so we choose to use the simpler method which requires only SSH (see Appendix B).

\subsection{Selecting for EF20191028}

We select the best and worst members using weekly RMSE of SSH compared against i) the verifying analysis and ii) using~\cite{cmemsdata} gridded SSH derived from satellite altimeter product.
A majority of the selected members (four of six) are shared between the benchmarks for both the best members and worst members, highlighting the robustness of the selection method. 
A possible reason for the minor difference in member selection may result from the verifying analysis containing observational information in addition to satellite-derived data.
Moreover, the horizontal resolution in the model grid is approximately a factor of {thirty} times finer than the {effective} satellite resolution~(\citealt{ballarotta2019resolutions}).
Despite this, the benchmarks agree well on which members do, and do not, capture SSH and the position of the LC. 

\begin{figure}[hbt!]
    \centering
    \includegraphics[width=0.7\linewidth]{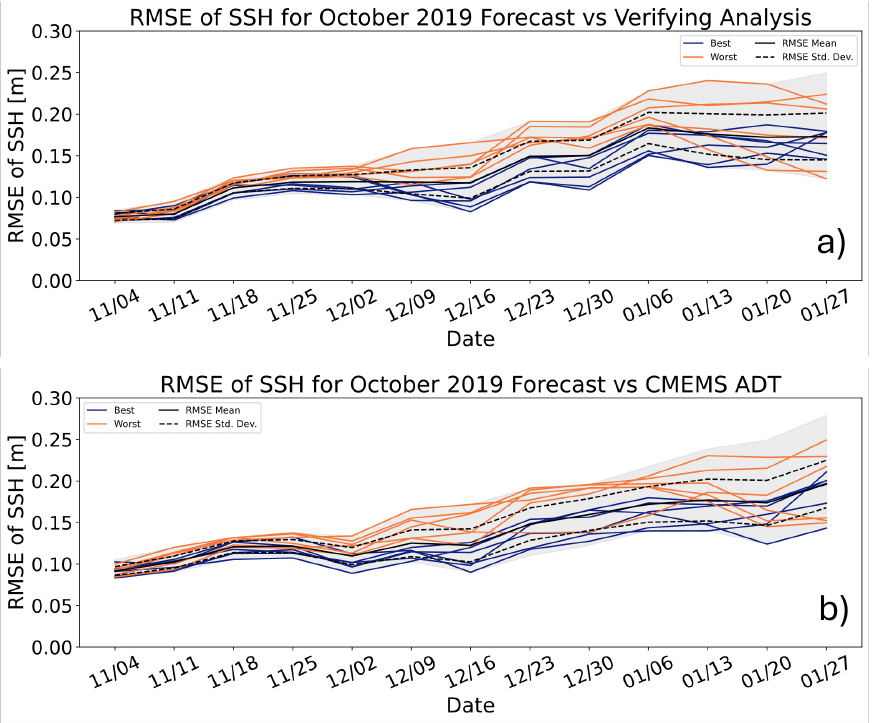}
    \caption{Weekly RMSE of SSH for EF20191028 compared against a) the verifying analysis and b) the CMEMS satellite product. Only the best (blue, solid lines) and worst (orange, solid lines) members are shown. The range (gray, shaded-region) is bounded by the weekly maximum and minimum values of RMSE. The mean of RMSE values {between all 32 members} (black, solid line) and one standard deviation above and below this mean (black, dashed lines) are provided.}
    \label{fig:EF2019_WklySSHRMSE}
\end{figure}

We plot the weekly RMSE of SSH for the two benchmarks in Fig.~\ref{fig:EF2019_WklySSHRMSE}.
For both, the spread of uncertainty is low over the first four weeks (11/04 to 11/25) with little distinction between the best and worst members.
After the fifth week (12/02), the two groups separate, with the best members maintaining consistently lower RMSE values over the next five weeks (through 01/06).
At the point of detachment for LCE Thor (01/13) and in the subsequent weeks, the best members maintain lower RMSE values of SSH compared against the mean. 
Remarkably, half of the members in the worst group improve, especially by the final week of the forecast (01/27), for both benchmarks.
For the ensemble mean, current forecast skill for SSH only extends for 5-6 weeks~(\citealt{thoppil2025evaluating}); hence, performance for the following weeks have higher uncertainty, especially in individual members. 
Despite this, we have confidence that the best (worst) members determined in the selection process are consistently the best (worst) over the majority of the forecast. 

\subsection{Selecting for EF20200106}

\begin{figure}[hbt!]
    \centering
    \includegraphics[width=0.7\linewidth]{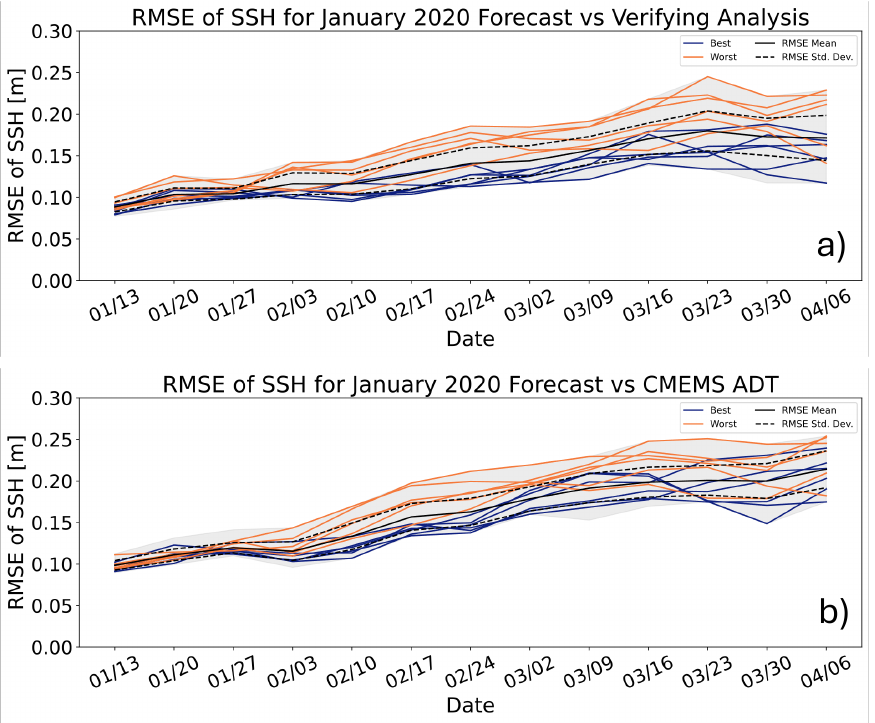}
    \caption{Weekly RMSE of SSH for EF20200106 compared against a) the verifying analysis and b) the CMEMS satellite product. Colors and line-types are the same as Fig.~\ref{fig:EF2019_WklySSHRMSE}}
    \label{fig:EF2020_WklySSHRMSE}
\end{figure}

We repeat the selection process for EF20200106, which includes the time-period of the reattachment and final separation of LCE Thor. 
As with EF20191028, the majority (four of six) of the best and the worst members are shared by the two benchmarks. 
This is especially encouraging due to the complex nature of the reattachment and separation occurring in this time-period.
Figure~\ref{fig:EF2020_WklySSHRMSE} displays the weekly values of SSH RMSE.
{We note} that the RMSE using the verifying analysis maintains better agreement -- lower maximum and minimum RMSE values -- over the forecast time-period, when compared to the CMEMS satellite product; however, the spread of the uncertainty is much larger, especially by the final week of the forecast. 
In contrast, the RMSE using the CMEMS satellite product has lower spread in the uncertainty and higher RMSE values, with a marked increase in slope of the RMSE occurring after the fifth week (02/10).
This behavior was not observed in the October 2019 forecast (Fig.~\ref{fig:EF2019_WklySSHRMSE}), as both benchmarks resulted in similar spread in uncertainty and maintained similar values over that time-period. 
We attribute this to the different stage in LCE separation process observed in each forecast. 

Much like the prior forecast, for both, the range of uncertainty is minimal over the first four weeks (01/13 through 02/03), and there is little differentiation between best and worst members.
For the verifying analysis at Week 5 (02/10), the uncertainty of the two groups becomes dissimilar, and between this week and shortly before the reattachment at Week 9 (03/09), the best (worst) members selected by the verifying analysis (Fig.~\ref{fig:EF2020_WklySSHRMSE}a) display consistently lower (higher) RMSE values than the mean of the RMSE (black, solid line).
Through the reattachment at Week 10 (03/16) to the final separation at Week 12 (03/30), a majority of the best members maintain their lower values of RMSE, compared against the mean of all RMSE values.
We note some overlap between individual members of the best and worst groups is observed in Fig.~\ref{fig:EF2020_WklySSHRMSE}a, but overall the worst members tend to display higher values of RSME during the forecast time-period. In contrast, the difference between the best and worst members is not as definitive with the CMEMS satellite product (Fig.~\ref{fig:EF2020_WklySSHRMSE}b). 
The two groups maintain distinction between Week 5 (02/10) and shortly before the reattachment (03/02); however, an increase in RMSE occurs for some of the best performing members with the opposite occurring for some of the worst performing members at the reattachment (03/09).
Despite this, by the time of the final separation of LCE Thor at Week 12 (03/30), the majority of the best members maintain lower uncertainty than the mean RMSE. 
Conversely, some of the worst members maintain RMSE values lower than the mean at this week. 
These results demonstrate the challenging nature in forecasting the reattachment and final separation of LCE Thor, and possibly indicates why the ensemble mean of the forecast performed so well during this time-period.
Although the uncertainty was larger when comparing the forecast against a satellite derived SSH, the spread of this uncertainty was lower, and we found that even those members that were initially doing poorly eventually improved.

\section{Structure of the Deep Eddy Field}

\subsection{EF20191028: Importance of Eddy Magnitude and Position}

\begin{figure}[hbt!]
    \centering
    \includegraphics[width=\linewidth]{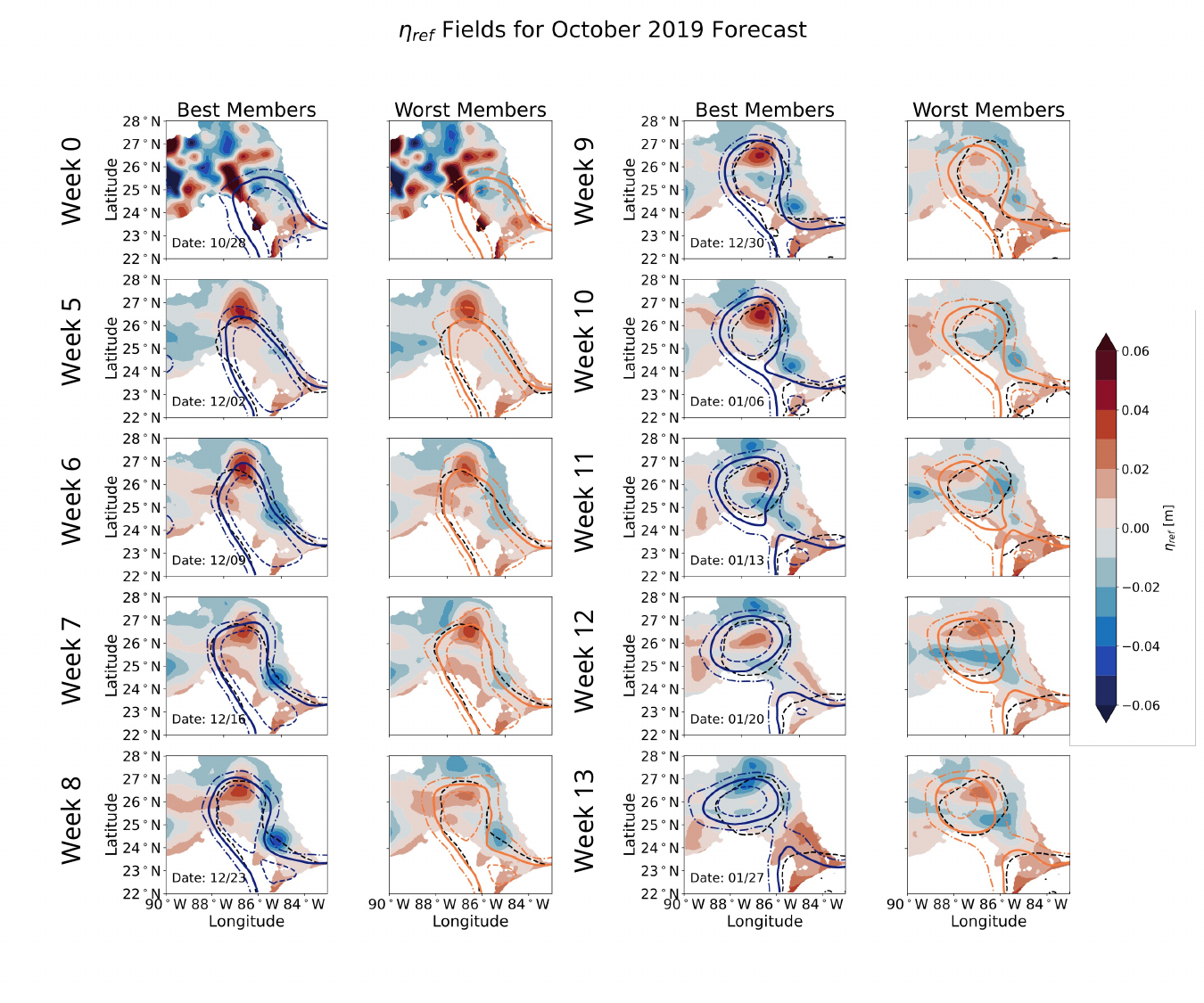}
    \caption{Mean $\eta_{ref}$ fields at 2000 m depth for the best (left columns) and worst (right columns) performing members for EF20191028 at Week 0 and Weeks 5-13.
    Deep cyclones are indicated with blue contours, and anticyclones with red; the upper is represented by three contour levels: the 17-cm contour (thick, solid line) which represents the LC, the $-3$-cm contour (dot-dashed line), and the 37-cm contour (dashed line) -- spanning the central two-thirds of the LC -- with the colors differentiating best members (blue) and worst members (orange) as in Fig.~\ref{fig:EF2019_WklySSHRMSE}. 
    For comparison, we include the 17-cm contour (black, dashed line) from the verifying analysis. }
    \label{fig:EF2019_upper_deep_comp}
\end{figure}

To better understand the  differences between the best and worst members, we review the deep field with a focus on the structure, location and propagation of mesoscale eddies.
Figure~\ref{fig:EF2019_upper_deep_comp} shows the mean upper and deep fields for the best and worst members. 
Weekly RMSE plots (Fig.~\ref{fig:EF2019_WklySSHRMSE}) indicated little difference in Weeks 1 to 4 in the upper field between best and worst members, so we show only the initial field (10/28) and the fifth (12/02) through thirteenth weeks (01/27).
For clarity, we highlight key findings:

\begin{itemize}
    \item The best members develop a strong anticyclone that persists over most of the forecast time-period, contributing to improved prediction of the LCE positioning;
    \item The best members show active development of cyclones off the Mississippi Fan, assisting in strengthening the DSC cyclone, as well as the eventual separation.
\end{itemize}

The initial deep fields (10/28) display nearly identical features, with minor deviations from the initial perturbations applied to the upper fields.
At Week 6 (12/09) a striking contrast develops between the mean $\eta_{ref}$ fields for the best and worst members.
Here, the best members show prominent, strong anticyclonic and cyclonic features: the anticyclone near the Mississippi Fan, centered at $26.5^\circ$N, as well as a developing cyclone beneath the Eastern leg of the LC (centered near $25^\circ$N), which is later observed to propagate towards the Deep Southeast Channel (DSC).
Both features are present in the worst member mean field; however, the magnitudes are weaker.
At Week 8 (12/23), the prominent features in the best members observed in Week 6 remain, whereas the worst members show an anticyclone with significantly decreased intensity, and a cyclone north of it ($27.5^\circ$N and $87^\circ$W).
More, the cyclone in the DSC (centered near $24^\circ$N and $85^\circ$W), although found in both groups, has greater magnitude in the best members.

In Weeks 10 through 12, the best member field retains the signature -- and most of the strength -- of the Mississippi Fan anticyclone.
The DSC cyclone for this group remains, possibly contributing to the initial detachment of Thor on Week 11 (01/13)~(\citealt{thoppil2025evaluating,safaie2025deep}).
The worst members retain a weaker DSC cyclone, aiding in predicting a detachment by Week 11 as well.
Despite this, the absence of the northern anticyclone results in an incorrect prediction of the positioning and orientation of the LCE, demonstrated by the misalignment of the 17-cm contour of the forecast and the verifying analysis (black, dashed line).
In contrast, the best member group displays a LCE with better predicted positioning and orientation.
Furthermore, in Weeks 11 to 13, we observe a northern cyclone developing off the Mississippi fan (centered at $27.5^\circ$N), which is not present in the worst member mean field.
These results suggest the worst members are missing deep mesoscale eddies, critical to predicting LCE positioning.
Overall, both member groups correctly predict the timing of the initial detachment of LCE Thor, as both maintain a cyclone in the DSC region, but we conjecture the stronger northern anticyclone and DSC cyclone resulted in better prediction of the location of the LCE after detachment for the best members.

\subsubsection{Comparison to Deep Observations}

\begin{figure}[hbt!]
    \centering
    \includegraphics[width=0.8\linewidth]{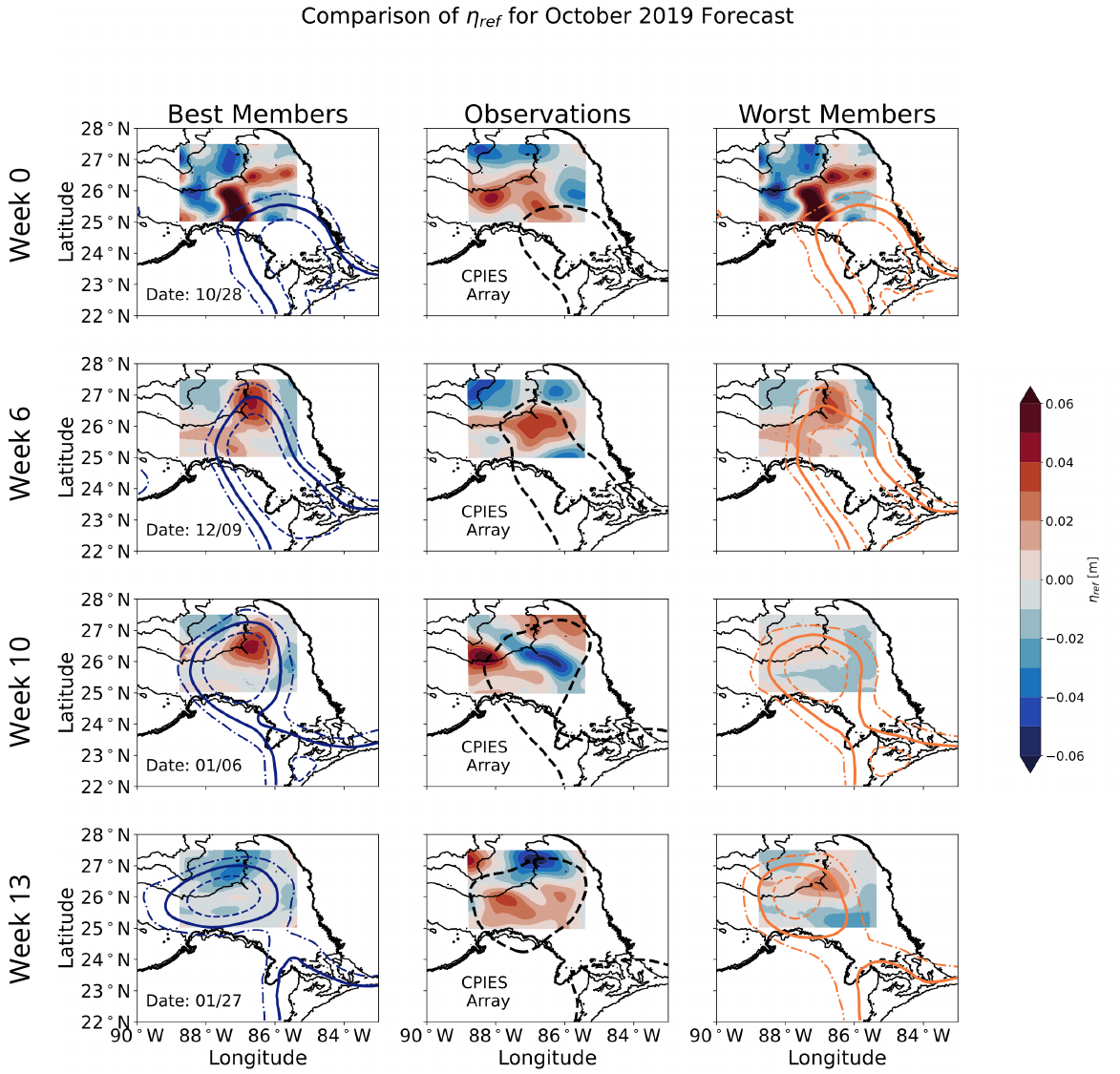}
    \caption{Weekly intervals of deep mean $\eta_{ref}$ field for the best performing members (left column), CPIES observation data (center column), and the worst performing members (right column) of EF20191028.
    The same colors and line-styles as Fig.~\ref{fig:EF2019_upper_deep_comp} are used.
    The models are restricted to the same region as the observations for direct comparison, and bathymetry is provided at 2000, 2500, and 3000 meters depth (solid, black contour lines).}
    \label{fig:EF2019_best_worst_obs}
\end{figure}

Figure~\ref{fig:EF2019_best_worst_obs} shows the mean $\eta_{ref}$ field for the best members (left column), the CPIES  (center column), and the worst members (right column).
 Model fields are limited to the same region as the observations for direct comparison, and we highlight the initial (Week 0) and final (Week 13) weeks of the forecast, as well as two weeks (Weeks 6 and 10) where the best members display better agreement with upper SSH fields (see Fig.~\ref{fig:EF2019_WklySSHRMSE}).

Qualitatively, the initial field {of} the model agrees with the observations{, as key deep circulation features are present in both. 
We emphasize this agreement is not to be expected as the model initial condition contains no information of the deep field.}
The model and observations both display a northern cyclone on the Mississippi Fan (near $27^\circ$N) with an anticyclone to the south.
At Week 6 (12/09), we find the northern anticyclone seen in the model is present in the observations with similar magnitude.
By Week 10 (01/06), the shape of the LC, based on the 17-cm contour is in better agreement for the best members.
The observations indicate the anticyclone is propagating west, with another anticyclone moving off the {Mississippi} Fan {near $27^\circ$N and $85.5^\circ$W) which will follow the bathymetry underneath the tip of the LC -- review of Week 9 and Week 11 confirm this, but are not shown here for brevity.}
By the final week of the forecast, the best members continue to show good agreement with the 17-cm contour from the satellite product, in terms of LCE position and orientation; moreover, the observations demonstrate a cyclonic eddy moving off the Mississippi Fan, which is also present in the best member mean field, and absent for the worst members. 
The observations indicate the model deep field, in the region of the CPIES array, initially agreed well, and it was the best members that continued to display similar, critical features resulting in better agreement in the upper field. 

\subsection{EF20200106: Predicting the Mississippi Fan Cyclone}

\begin{figure}[bht!]
    \centering
    \includegraphics[width=\linewidth]{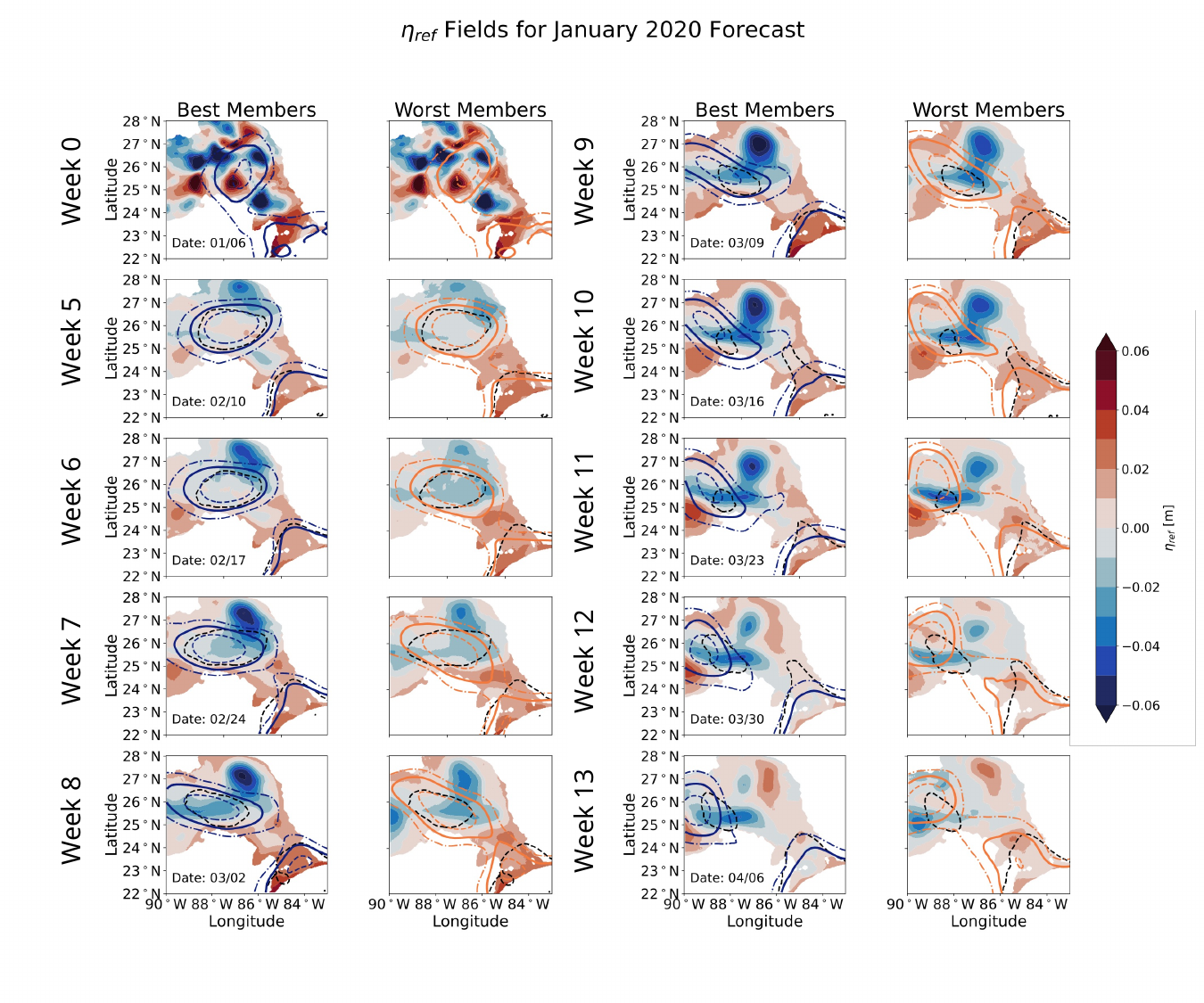}
    \caption{Weekly intervals of mean $\eta_{ref}$ field for the best (left columns) and worst (right columns) performing members of EF20200106. 
    Colors and line styles are the same as Fig.~\ref{fig:EF2019_upper_deep_comp}.
    }
    \label{fig:EF2020_upper_deep_comp}
\end{figure}

Mean $\eta_{ref}$ fields for EF20200106  are given in Fig.~\ref{fig:EF2020_upper_deep_comp}.
As before, we briefly summarize key findings that are explained in greater detail below:

\begin{itemize}
    \item Both groups develop a cyclone off the Mississippi Fan;
    \item Best members develop a stronger cyclone earlier in the forecast contributing to improved prediction in the separation and positioning of LCE Thor;
    \item Worst members develop a strong anticyclone in the DSC, resulting in a misaligned LC and the attachment of the $-3$-cm contour.
\end{itemize}

Initial fields are nearly identical, as before, with prominent differences appearing by Week 5 (02/10).
Now, the best member mean field displays a strong, northern cyclonic feature (centered near $27.5^\circ$N and $87^\circ$W) that is also present in the worst member mean field, but with lower magnitude and centered more south and east ($26.5^\circ$N and $86^\circ$W).
Further, the best member field displays a cyclonic feature between the LCE and the LC that remains until Week 6 (02/17).
In contrast, the worst member field shows an anticyclonic region south of $25^\circ$N.
Between Week 5 (02/10) and Week 9 (03/09), for the best members we observe the northern cyclone moving south off the Mississippi Fan, and growing stronger.
Comparatively, the worst member northern cyclone peaks in magnitude as LCE Thor reattaches (03/16), but weakens over the following weeks (03/23 to 04/06). 
At this point, an additional cyclonic region appears underneath the LCE.  
By the time of the reattachment (03/16), both groups maintain a cyclone located near $25.5^\circ$N and $87.5^\circ$W with similar magnitude, as well as a cyclonic region under the LCE.
At Week 12 (03/30), the strength of these features is much greater for the best members, and we observe the cyclone constrained by the Campeche Bank and Mississippi Fan remains stationary through the end of the forecast with consistent strength, whereas it propagates west, losing intensity, for the worst members.

Focusing on the DSC region, at Week 11 we observe a stronger anticyclone at $24^\circ$N and $86.5^\circ$W for the worst members, which persists until the end of the forecast.
For the best members, this feature is absent, only appearing near Week 13.
These differences are reflected in the upper field by the positioning of both the path of the LC and the LCE itself.
Here, the best members show separation at all SSH contours ($-3$-cm, $17$-cm, and $37$-cm), as well as a LC in a port-to-port configuration; this better resembles the $17$-cm contour of the verifying analysis.
In contrast, the $-3$-cm contour for the worst members remains attached to the LC, and the LC extends as far north and west as $24.5^\circ$N and $87^\circ$W, most likely due to the presence of the deep anticyclone.

\subsubsection{Comparison to Deep Observations}

\begin{figure}[hbt!]
    \centering
    \includegraphics[width=0.8\linewidth]{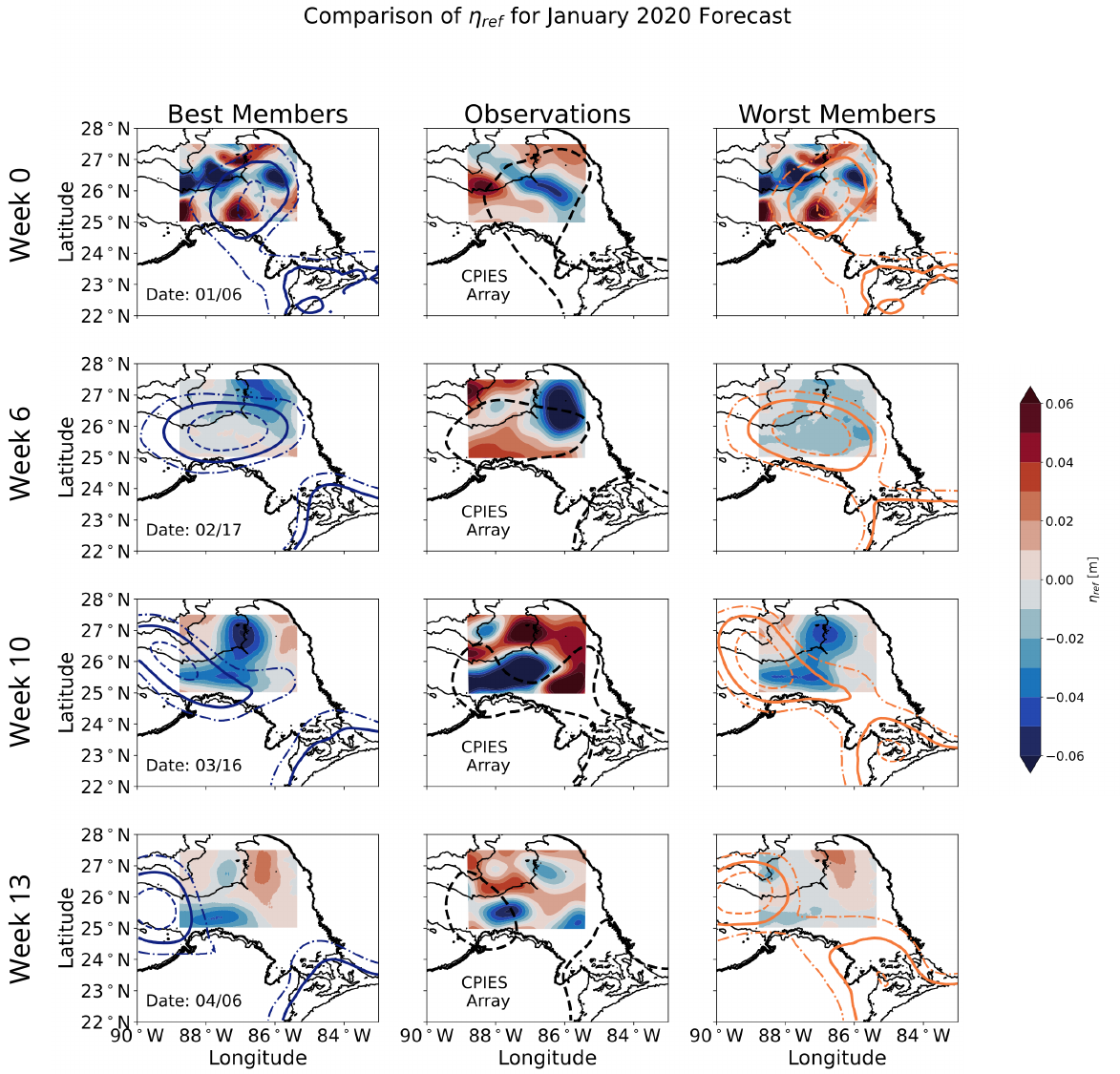}
    \caption{
    Weekly intervals of deep mean $\eta_{ref}$ field for the best performing members (left columns), CPIES observation data (center columns), and the worst performing members (right columns) of EF20200106.
    We use the same colors, line-styles, and bathymetry as Fig.~\ref{fig:EF2020_upper_deep_comp}.
    The models are restricted to the same region as the observations for direct comparison.
    }
    \label{fig:ef2020_best_worst_obs}
\end{figure}

We compare the best and worst member mean fields of $\eta_{ref}$ to the CPIES array (Fig.~\ref{fig:ef2020_best_worst_obs}).
The initial (Week 0) fields from the model are found to be in poor agreement compared to the observations, in contrast to the prior forecast. 
The CPIES array shows a westward anticyclone coupled with an eastern cyclone in the observation region, whereas the model displays cyclonic features in the east and west of the same region, flanked by anticyclonic features to the north and south. 
At Week 6 (02/17), model magnitudes are significantly lower than the observations.
However, both model groups contain a cyclonic feature near the Mississippi Fan (near $27^\circ$N), resembling the strong cyclonic feature observed by the CPIES array. 
Best members show better agreement in orientation and positioning of the LC and LCE.
Near the reattachment at Week 10 (03/16), the observations indicate the deep cyclone has propagated south, and is constrained by the bathymetry between the Mississippi Fan and the Campeche Bank (near $25.5^\circ$N).
For the model, two cyclonic features are present: i) a strong cyclone off the Mississippi Fan; and ii) a cyclone between the Mississippi Fan and the Campeche Bank (near $25.5^\circ$N).
For the more northern cyclone, the best members display a feature of stronger magnitude, resulting in a deeper trough in the $-3$-cm contour in contrast to the worst members; however, both fields have weak agreement with the satellite product.
By Week 13, the southern, constrained cyclonic feature persists for the best members in a similar position as that observed by the CPIES (found near $25.5^\circ$N and $87.5^\circ$W), which is {further west} for the worst members. 
This aids in the best members showing somewhat better agreement with the 17-cm contour from the satellite product. 

Although the initial field was qualitatively different from the observations, both the best and worst member model fields developed the cyclone seen in the observations near the Mississippi Fan on 02/17.
This is most likely due to interactions of the LCE and the topography, leading to baroclinic instabilities~(\citealt{donohue2016loop}). 
The timing of the cyclone's development in the model was delayed, leading to an initially weaker, comparatively more northern cyclone, with the worst members showing a more pronounced delay in the development of the cyclonic feature.
The model, in both groups, recovers, with the additional cyclonic region forming beneath the LCE.
The presence of the strong Mississippi {Fan} cyclone, along with the additional cyclonic region, aids in the reattachment and final separation of LCE Thor in the model. 
The time-delay and reduced model strength of the northern cyclone most likely led to the contrast between the model SSH field and the verifying analysis (and observations).
Further, the poor initialization of the deep field played a role in the delay of the Fan cyclone development, experienced by both the best and worst members; proper initialization of the deep could have prevented such delay. 
More, this likely resulted in the increased range of uncertainty in SSH, seen in Fig.~\ref{fig:EF2020_WklySSHRMSE}.

\section{Summary and Implications for Future Forecasts}

Deep ($> 1000$ m depth) mesoscale eddies play an important role in the performance of ensemble forecast members in predicting sea surface height (SSH) in the Loop Current (LC) system in the Gulf of Mexico.
{This work set out to (i) identify the influence of the deep initial field on the ensemble forecast, and (ii) review how the deep-field predictions varied within an ensemble forecast.}
Here, 92-day, 32-member ensemble forecasts that span the Loop Current Eddy (LCE) Thor separation event, from October 28th, 2019 through April 6th, 2020 are used. 
We first identified that smaller uncertainty in the initial deep fields led to an improved SSH forecast.
Two forecasts{, chosen due to their success in capturing the initial detachment, reattachment, and then final separation of LCE Thor,} provided case studies.
With these two forecasts, a simple, yet effective method to rank the performance of individual ensemble forecast members was developed using root-mean-square-error (RMSE) of SSH against the verifying analysis, as well as SSH from {a} CMEMS satellite product (Fig.~\ref{fig:memSelect}).
With the calculated RMSE values, the best $80^{th}$ and worst $20^{th}$ percentiles of ensemble members over the 13 week forecast time-period were found.
The two groups were found to {become} distinct near Week 5 of the forecast, with the best members typically maintaining lower uncertainty over the forecast time-period (Figs.~\ref{fig:EF2019_WklySSHRMSE} and~\ref{fig:EF2020_WklySSHRMSE}).
Further, the verifying analysis and CMEMS satellite product are both capable of identifying the best and worst members, with both benchmarks indicating the same majority of members.

The deep field was represented by $\eta_{ref}$, the $2000$ m streamfunction.
In both forecasts the deep circulation displayed distinct differences between best and worst members, particularly in the magnitudes and positioning of deep cyclones and anticyclones.
Through comparison with deep observations collected with a CPIES array,  best members were shown to have better{, although marginally so,} agreement with observed prominent deep features (Figs.~\ref{fig:EF2019_best_worst_obs} and~\ref{fig:ef2020_best_worst_obs}). 
{Presently, these ensemble forecasts have no predictive skill, as no deep ocean observations are assimilated.}
Moreover, when the initial deep field did not agree with observations, the spread of model ocean state realizations increased. 
{These comparisons highlight an opportunity for improving forecasts through assimilation of deep observations.}

Future forecasts in the Gulf, and more broadly in other boundary current systems, should include deep ocean observations in their assimilation schemes to improve initial {upper and deep} ocean states.
{Alternatively, using temperature and salinity profiles from the CPIES would improve the upper ocean initial state, with the desire that the deep field adjusts accordingly.}
From the two ensemble forecasts reviewed in the present study, a more realistic deep ocean eddy field has a positive influence on the forecast. 
In the first case study, the best members displayed better agreement with deep observations (see Figs.~\ref{fig:IC_vs_wk6} and~\ref{fig:EF2019_best_worst_obs}).
We postulate this initialization in the deep led all ensemble members to better forecast the initial detachment of LCE Thor. 
However, differences in the evolution of the mean field between the two led to better prediction of SSH in the best member group.
In our second case study, the initial deep mean $\eta_{ref}$ field disagreed with observations (see Figs.~\ref{fig:IC_vs_wk6} and~\ref{fig:ef2020_best_worst_obs}).
This difference likely led to greater uncertainty in the ensemble, particularly for the worst member group (Fig.~\ref{fig:EF2020_WklySSHRMSE}). 

This study reviewed the Thor LCE separation, thus it would be prudent to review other separations with forecasts that have successfully, or unsuccessfully, predicted LCE separation cycles.
In addition, these results indicate that deep circulation matters, but our understanding of when and where these observations are most critical for forecast success is not well known. 
Another study could follow~\cite{dukhovskoy2023assessment} and conduct an Observing System Experiment (OSE) and an Observing System Simulation Experiment (OSSE). 
{An OSE would assess the impact of including existing deep observations -- such as bottom pressure and near-bottom currents which constrain the depth-independent mode -- from the CPIES on predicting the LCE separation process.}
For a future OSSE study, it would be beneficial to include an investigation of the role of {deep} cyclones{, through inclusion of synthetic CPIES observations} in the Deep Southeast Channel (DSC) region, which has been shown to influence LCE separation events (\citealt{thoppil2025evaluating,safaie2025deep}).

\clearpage
\acknowledgments
J.P.C., K.A.D., and D.R.W. acknowledge funding from the US National Academy of Science, Engineering, and Medicine (NASEM) under their Gulf {Research} Program, UGOS grants, 2000009943 and 200013145. 
J.P.C., K.A.D., and D.R.W. also acknowledge funding from the US Naval Research Laboratory -- Stennis Space Center \textit{via} Office of Naval Research Cooperative Agreement N00173-22-2C005. 
P.G.T. acknowledges support from the CASE EJIP funding from Chevron Research and Technology.


%
%
\datastatement
The CPIES datasets analyzed for this study can be found in the GRIIDC data repository: \url{https://data.griidc.org/data/U1.x852.000:0004} and DOI:10.7266/BZ9B3C54.

%



\appendix[A] 

\appendixtitle{Surface Velocity as an Additional Metric}

Surface velocity is another parameter which is useful to identify the LC and LCEs, as the 1.5-kt contour is typically taken to represent its frontal edge~(\citealt{ivanov2024process}).
Here, we test the use of the surface velocity as an additional measure of model performance{. 
We compare the RMSE of surface velocity magnitude, taken from the verifying analysis, against the RMSE of SSH also taken from the verifying analysis. 
This ensures we compare at the same points and that the expected linear relationship exists in the model.}
In both the model and the analysis, we determine the magnitude of the velocity using the longitudinal and meridional components at each surface point, converting from units of meters-per-second to knots before conducting the calculation. 

\begin{figure}[hbt!]
    \centering
    \includegraphics[width=0.75\linewidth]{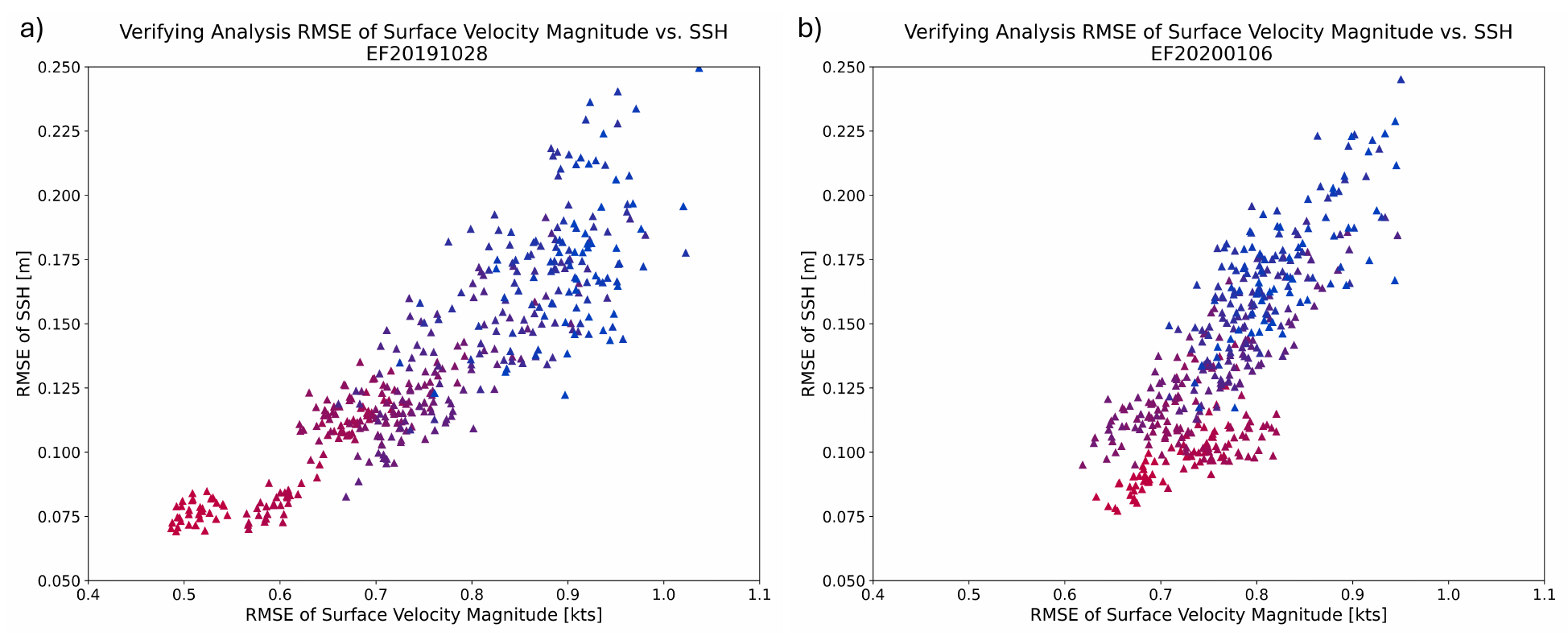}
    \caption{Scatter plot of weekly RMSE of surface velocity magnitude versus RMSE of SSH for all forecast members of a) EF20191028 and b) EF20200106. Colors correspond to the week from forecast start time, beginning from Week 1 in pink to Week 13 in blue.}
    \label{fig:ssh_vs_vel}
\end{figure}

We directly investigate the relationship between RMSE of SSH and RMSE of velocity; we plot this for each individual member of the two forecasts, to identify the link between the two parameters and individual member performance. 
For both forecasts, we observe a linear relationship between RMSE of SSH and RMSE of velocity, indicating that ensemble members that perform well in one metric generally do well in the other. 
Further, the forecast uncertainty increases with time, as lower, more concentrated RMSE values are found for Week 1 (pink triangles), whereas Week 13 (blue triangles) shows both high RMSE and large spread of possible values. 
Both surface metrics perform favorably, so we choose to use SSH in our selection process of best and worst members.

\appendix[B]

\appendixtitle{Weighted RMSE Selection}

We examined a weighted RMSE metric which used SSH, as well as $T$, $S$, $U$, and $V$ from the upper $500$ meters of the water column.
{This method was developed to provide a comprehensive test of surface variables; 
however, through geostrophy and thermal wind there should be close correspondence between these variables at these scales. 
Hence, we expect this method to select similar members as that using only SSH.}
Weekly RMSE values were calculated between each ensemble member of the forecast and the verifying analysis at each depth.
A weighted RMSE ($\hat{X}$) was calculated using:

\begin{equation}
    \hat{X}(z) = 
    \begin{cases}
        0.25\text{RMSE}(SSH) + 0.25\text{RMSE}(T) + 0.25\text{RMSE}(S) + 0.125\text{RMSE}(U) + 0.125\text{RMSE}(V), \text{ } z = 0\\
        0.375\text{RMSE}(T) + 0.375\text{RMSE}(S) + 0.125\text{RMSE}(U) + 0.125\text{RMSE}(V), \text{ } z < 0
    \end{cases} 
\end{equation}

\noindent Using $\hat{X}$ found at each depth, members were ranked using the method outlined in Fig.~\ref{fig:memSelect}.
{For the variables constrained through DA (SSH, $T$, and $S$), higher weights are assigned.
As expected}, this {method }yielded a majority of the same members selected as the best and worst for both forecasts compared against the members selected using RMSE of SSH -- for both the verifying analysis and the satellite product from CMEMS. 
{We carry out our analysis with} the simpler method that only requires use of SSH.

\appendix[C]

\appendixtitle{{Details of the Ensemble Forecasts}}

For completeness, we briefly summarize the model details from~\cite{thoppil2025evaluating}.
The ensemble forecast system used in the present study~(\citealt{thoppil2025evaluating}) was developed at the Naval Research Lab, and combines the Navy Coastal Ocean Model (NCOM;~\citealt{martin2009user}), the Navy Coupled Ocean Data Assimilation System (NCODA;~\citealt{cummings2005operational}), and the Coupled Ocean-Atmosphere Mesoscale Prediction System (COAMPS;~\citealt{hodur1997naval}). 
The NCODA system handles the data assimilation for the forecasts, deploying a 3-D Variational (3DVar) technique to assimilate observations in near-real time. 
These observations include satellite altimeter tracks of sea-surface height (SSH), sea-surface temperature (SST), as well as temperature and salinity profiles from ships, gliders, and floats. 
Surface observations of SSH and SST are input into the Modular Ocean Data Assimilation System (MODAS) to construct synthetic profiles of temperature and salinity, which are extended down to $1000$ meters to constrain the ocean interior~(\citealt{fox2002modular}).
At the time of the forecasts, assimilation of velocity observations was not possible, thus they are not included. 

\subsection{Model Initialization}

To run the forecast requires conducting two separate model runs. 
A control run, which represents a single deterministic forecast, is used to generate the perturbations for the model initial conditions with the 24-hour forecast error variances and the Ensemble Transform (ET) methodology~(\citealt{bishop1999ensemble}). 
Additionally, the initial condition perturbations are supplemented with additional estimates of model temporal variability, nowcast/analysis increments history, and climate variability such that the ensemble model perturbations will have a spread similar to the best guess of the control run analysis error variance. 
This is critical to ensuring the distribution of the ensemble solutions can include all detectable and dynamically relevant ocean states. 

Surface boundary conditions from COAMPS -- wind and heat fluxes -- are also perturbed during the forecast using a space-time deformation technique with random shifting~(\citealt{wei2016performance}). 
Doing so addresses uncertainties in surface forcing by emulating displacement and time lag errors in dominant dynamical atmospheric forcing features through random temporal shifts.
For individual ensemble members, the same 3-hour interval is used to prepare the forcing, but the values are computed at randomly shifted times through the linear interpolation of the original forcing fields. 
To ensure interpolated fields are uncorrelated beyond a 24-hour interval, every 24 hours, independently generated random fields are used to derive the time shifts with a specified spatial de-correlation. 
This results in independent forcing for each ensemble member~(\citealt{coelho2009note,wei2016performance}).

\subsection{Details of the Boundary Conditions}

For the surface boundary conditions, COAMPS is used.
COAMPS has a forecast length of 5 days, so for extended range forecasts, such as those used herein, climatological forcing is constructed for the lateral and surface boundary conditions. 
To generate the 3-hour fields for the annual climatology, 8 years of 3-hourly output (2005-2013) from the Central America COAMPS are used. 
To handle the lateral boundary conditions a similar method is employed for the lateral forcing.
Daily climatology from 12-years (1994-2005) of output from a global HYCOM reanalysis was constructed~(\citealt{metzger2014us,thoppil2016current}).
For these forecasts, the first 5-days were from a global HYCOM, then transitioned to the daily climatology.

%



\bibliographystyle{ametsocV6}
\bibliography{main}

\end{document}